\documentclass{article}
\usepackage{spconf,amsmath,graphicx}
\usepackage{xspace}

\usepackage{enumitem}

\usepackage{times}
\usepackage{textcomp}

\usepackage{varwidth}

\usepackage{tabu}

\usepackage{nicefrac}

\usepackage{mwe}

\usepackage{amsmath, lipsum}

\usepackage{tabularx}
\usepackage{booktabs}
\newcolumntype{R}{>{\centering\arraybackslash}X}
\usepackage{multirow}
\newcolumntype{L}{>{\centering\arraybackslash}m{3cm}}
\usepackage{adjustbox}
\newcolumntype{Y}{>{\centering\arraybackslash}X}

\usepackage{mathtools}
 
\usepackage{blindtext}
\usepackage{graphicx}

\usepackage{amsmath}
\usepackage{algorithm}
\usepackage[noend]{algpseudocode}
\makeatletter
\def\BState{\State\hskip-\ALG@thistlm}
\makeatother

\algnewcommand\algorithmicforeach{\textbf{for each}}
\algdef{S}[FOR]{ForEach}[1]{\algorithmicforeach\ #1\ \algorithmicdo}

\renewcommand\algorithmicthen{}

\algnewcommand{\IfThenElse}[3]{
  \State \algorithmicif\ #1\ \algorithmicthen\ #2\ \algorithmicelse\ #3}

\DeclareMathOperator*{\argmin}{arg\,min}

\algrenewcommand\textproc{}
 
\usepackage{lettrine}

\usepackage{tabularx}
\usepackage{booktabs}

 \usepackage{subfig}
\usepackage{tabularx}

\usepackage{url}

\usepackage{enumitem}

\usepackage{cite}

\usepackage{adjustbox}
\newcolumntype{Y}{>{\centering\arraybackslash}X}

\newcolumntype{?}{!{\vrule width 1.5pt}}

\usepackage{diagbox}

\usepackage{mathtools}

\newcommand\numberthis{\addtocounter{equation}{1}\tag{\theequation}}

\title{Deep Selector-JPEG: Adaptive JPEG Image Compression for Computer Vision in Image classification with Human Vision Criteria}

\name{Hossam~Amer, Sepideh Shaterian, and~En-hui~Yang, \textit{Fellow, IEEE}\thanks{THIS WORK WAS SUPPORTED IN PART BY THE NATURAL SCIENCES
AND ENGINEERING RESEARCH COUNCIL OF CANADA UNDER
GRANT RGPIN203035-16, AND BY THE CANADA RESEARCH
CHAIRS PROGRAM.}}

\address{University of Waterloo,
	Electrical and Computer Engineering,
	Waterloo, ON, Canada N2L 3G1}
 
\begin{document}
%
\maketitle
\begin{abstract}

With limited storage/bandwidth resources, input images to Computer Vision (CV) applications that use Deep Neural Networks (DNNs) are often encoded with JPEG that is tailored to Human Vision (HV). This paper presents Deep Selector-JPEG, an adaptive JPEG compression method that targets image classification while satisfying HV criteria. For each image, Deep Selector-JPEG selects adaptively a Quality Factor (QF) to compress the image so that a good trade-off between the Compression Ratio (CR) and DNN classifier Accuracy (Rate-Accuracy performance) can be achieved over a set of images for a variety of DNN classifiers while the MS-SSIM of such compressed image is greater than a threshold value predetermined by HV with a high probability. Deep Selector-JPEG is designed via light-weighted or heavy-weighted selector architectures. Experimental results show that in comparison with JPEG at the same CR, Deep Selector-JPEG achieves better Rate-Accuracy performance over the ImageNet validation set for all tested DNN classifiers with gains in classification accuracy between 0.2\% and 1\% at the same CRs while satisfying HV constraints. Deep Selector-JPEG can also roughly provide the original classification accuracy at higher CRs.

\end{abstract}


\section{Introduction}
\label{sec:intro}

Almost daily there are large amounts of images that needs to either be stored for or exchanged among Computer Vision (CV) applications. Deep learning (DL) is a key to these CV applications due to its ability to extract desired features from raw pixels of input images without any domain knowledge \cite{hinton_2015_edges}. To extract these features in the task of image classification, for instance, deep neural networks (DNNs) learn the parameters of non-linear activation functions using a backpropagation learning algorithm. These functions progressively transform raw pixels of the input image to produce the output predicted label \cite{hinton_2015_edges}. With this capability, DL showed success in image classification with a steady accuracy improvement on the large-scale and high-quality ImageNet dataset from 63.3\% to 90.2\% \cite{hinton_2015_edges, pham2020meta}.

Raw pixels of these large-scale image datasets fed to underlying DNNs typically come from the pipeline of image acquisition, encoding, storage/transmission, and decoding. This implies that these raw pixels are indeed compressed in a lossy manner to meet the storage and bandwidth requirements. 

Since the late 1980s, JPEG codec has been a widely used codec for images to control the trade-off between compression rate and human perceived quality via a parameter called Quality Factor (QF). Yet, JPEG paid little attention to CV \cite{jpeg_standard, yang2008joint}. We have conducted an experiment that shows that if QF = 10 is used to compress all images in the ImageNet validation set, a Compression Ratio (CR) of 11x can be achieved at the expense of a drop of $\approx$8-10\% in terms of classification accuracy of DNN classifiers. Even if the image perceptual quality at QF=10 is deemed acceptable according to Human Vision (HV), the $\approx$8-10\% drop in classification accuracy may be too significant to be absorbed for CV. Along the same lines, Table \ref{tab:bird651} shows an image instance from the ImageNet validation set fed to the Inception V3 (IV3) model at different quality factors. This table indicates that a JPEG compressed version with a lower QF could yield a higher rank of the ground-truth (GT) label in comparison with the original image with a reasonable HV quality \cite{amer2020adaptive, amer2021compression}. Therefore, it would be desirable to adaptively select the QF of each input image of the large-scale image dataset to improve the trade-off between the JPEG CR and DNN classification Accuracy (Rate-Accuracy (RA) performance) while maintaining certain perceptual quality for humans. The question is, of course, how?

\begin{table}[htbp]
\centering \normalsize
\caption{Image ID\#651 From ImageNet Fed to Inception V3.}
  \begin{tabular}{ | c | c | c | c | c | }
    \hline
     &  Original & QF=60 & QF=40 & QF=10 \\
    \hline
     PSNR & Inf & 35.2 & 33.5 & 29.9 \\  \hline 
     MS-SSIM & 1.0 & 0.98 & 0.97 & 0.95 \\  \hline 
    Rank for GT Label & 2 & 2 & 2 & 1 \\  \hline 
  \end{tabular}
  \label{tab:bird651}
\end{table}

In the literature, two classes of methods were used to answer the above question \cite{shen_unified_features_vcm_2018, hu2020towards, liu2018deepn, xie2019source}. The first class successfully used neural networks trained end-to-end jointly with HV and CV or only HV constraints \cite{balle2018variational, minnen2018joint, theis2017lossy, toderici2017full, minnen2017spatially, johnston2018improved, hu2020towards, chamain2021end}. Some papers of the latter set targeting both HV and CV retrained jointly the DNN used for CV tasks along with the neural networks used for compression. The DNN used for CV tasks may require large training resources or not always be known, which makes joint training challenging. Also, their methods were not evaluated on several DNNs for the task of image classification. The second class redesigned quantization tables of classical codecs to suit HV and different CV tasks \cite{liu2018deepn, xie2019source, brummer2020adapting, ghamsarian2020relevance}. However, not all of these papers adaptively selected JPEG's QF of each input image for image classification while satisfying HV criteria and testing on the entire ImageNet under different DNNs.

This paper presents Deep Selector-JPEG, an adaptive JPEG compression method that uses DNN classifiers to target CV in image classification while meeting HV criteria. For each image in the training set, Deep Selector-JPEG labels the set of feasible QFs to compress an image. The set of feasible QFs is determined based on two criteria: (1) MS-SSIM value of the JPEG compressed image with any feasible QF is greater than or equal to the MS-SSIM threshold with a high probability, where the probability is calculated as if the image is taken randomly and uniformly from an image set, say, ImageNet validation set; (2) Rank for the GT label of the JPEG compressed image with the given QF is either the same or better than the rank of the GT label of the original image in ImageNet. For each given QF, we train one independent binary DNN classifier to predict whether this given QF is feasible. Deep Selector-JPEG is designed via light-weighted or heavy-weighted selector architectures. At inference time, Deep Selector-JPEG selects the least feasible QF to compress the original image.

Experimental results show that in comparison with JPEG at the same CR across $10$ different image classification test DNNs, Deep Selector-JPEG architecture achieves better RA performance over the entire ImageNet validation set for all tested DNN classifiers with gains in the classification accuracy range between $\approx$0.2\% and $\approx$1\% at the same CRs while satisfying HV constraints. Deep Selector-JPEG can maintain the original classification accuracy at higher CRs

\section{Deep Selector-JPEG: Adaptive JPEG Compression For Image Classification with Human Vision Criteria}
\label{sec:selector_jpeg}

\subsection{Deep Selector-JPEG: Formation and Considering Image Classification With Human Vision Criteria}

Deep Selector-JPEG targets image classification while satisfying human vision criteria by the adaptive selection of a QF for each input image from a set of feasible QFs. Let $L=\{10, 20, 30, 40, 50, 60, 70, 80, 90\}$\footnote{This set of QF values is simply used as an example. The idea of this paper, however, can be applied to any set of QF values. In addition, $\mbox{QF}=10$ is regarded in this example as the lowest compression quality acceptable to humans.}. A QF is feasible in $L$ if it satisfies two constraints:

\begin{enumerate}[topsep=0pt,itemsep=-1ex,partopsep=1ex,parsep=1ex]
    \item MS-SSIM of the compressed image with this QF is greater than a target MS-SSIM predetermined by humans with a high probability \cite{kostina2012fixed}.     
    \item Rank of the GT label of the compressed image with this QF is either the same or better than the rank of the original image in the ImageNet set.
\end{enumerate}

Based on the above constraints, our training set is $T = \{(x_1, q_1), \cdots, (x_N, q_N)\}$, where $X$ is the set of all original images in the ImageNet training set, and for each original image $x_i$,  $q_i$ is a binary vector $q_i = (q_{i,1}, \cdots, q_{i,n})$ with  $q_{i, j} = 1$ indicating that $QF_j$ is feasible for the original image $x_i$, where $X$ denotes the set of all original images.

To satisfy the first feasibility constraint, we offline created a cluster of compressed images for each $QF$ $\in$ $L$ from the original ImagetNet images. For a target MS-SSIM, we calculated the percentage of images, i.e the hit rate, in each QF cluster with MS-SSIM greater than or equal to the target MS-SSIM. Then, we remove QFs in $L$ whose hit rate is less than 90\%. As for the second constraint, the ground truth label $q_{i, j}$ ideally should be determined by humans. That is, given the original image $x_i$ and its JPEG compressed image with $QF_j$, a human should determine whether the compressed image would lead the human to believe that the rank of GT label is still at least maintained (i.e feasible) with respect to its original image. Because of the sheer size of the image set, such a task is daunting. To alleviate this difficulty, we instead replace human labelers with a pre-trained DNN classifier $S$. 

\begin{figure} [htbp]
\setlength{\belowcaptionskip}{-4ex}
      \centering
             \includegraphics[width=1\linewidth]{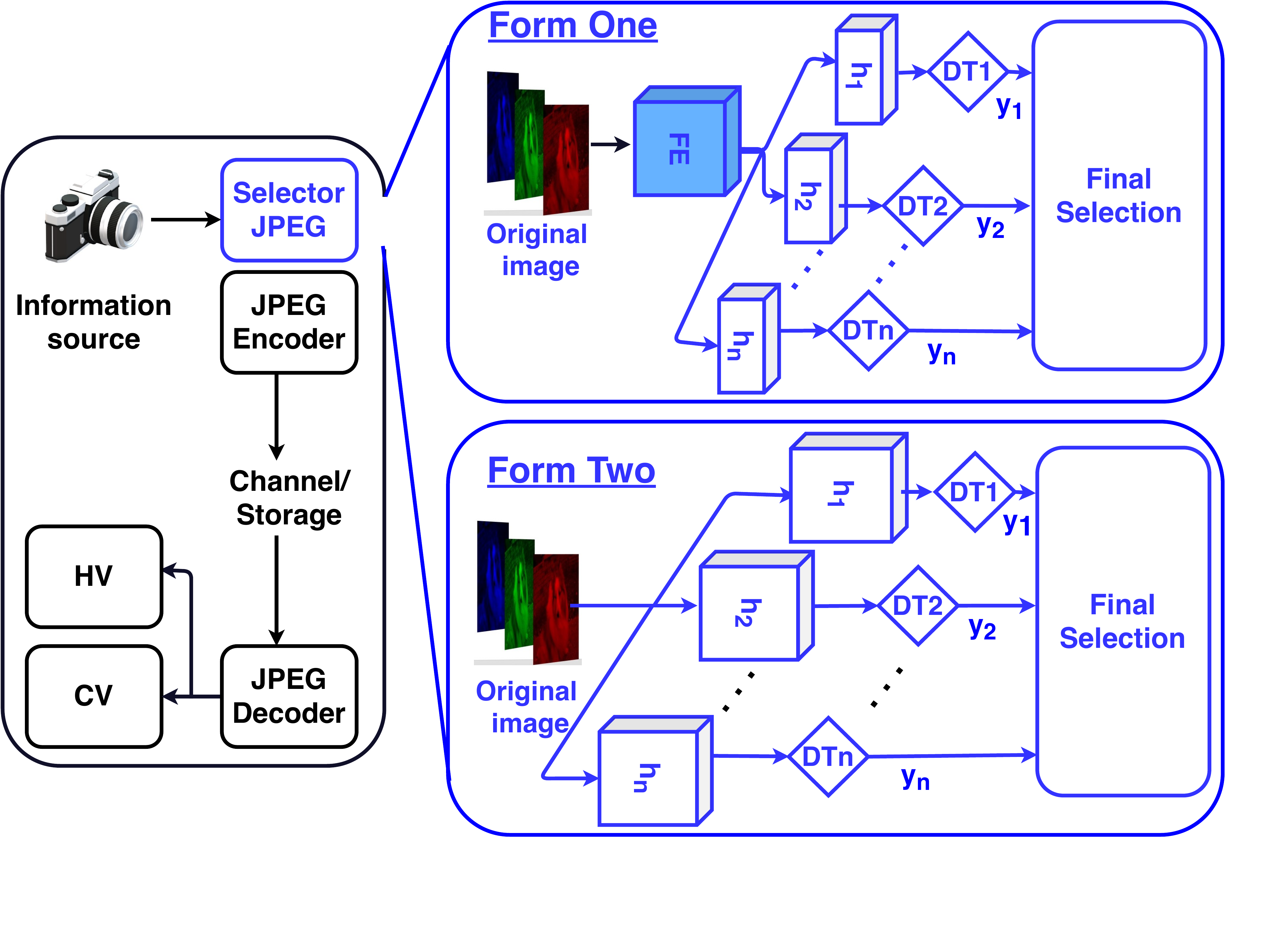}
      \caption{Deep Selector-JPEG Formation and Architecture.}
      \label{fig:selector_jpeg_hossam_sep}
\end{figure}

In Figure \ref{fig:selector_jpeg_hossam_sep}, Deep Selector-JPEG determines each hypothesis $y_j$: $X\xrightarrow[]{}\{0, 1\}$  via ``deep'' supervised learning, using two forms of  DNN architectures. In both forms, we decompose the given problem with $n$ QFs into $n$ independent binary classification problems. Thus, each hypothesis $y_j$ is induced to predict the feasibility of its corresponding $QF_{j}$, $j=1, 2, \cdots, n$, inside $L$ for each original image $x \in X$. In terms of architecture, Form One freezes a pre-trained image classification DNN $S'$ except for its last two blocks and uses the frozen section as a common feature extractor for all binary classifiers. Based on the common features, each $h_{j}$ that consists of the last two blocks of $S'$ is induced during training to make its own independent binary classification decision, $y_{j}$. To further improve RA performance, the entire $S'$ is learned for each $y_{j}$ independently in Form Two.

For each binary classifier, the following binary cross-entropy loss function is utilized to obtain optimized weights $W_j^*$:

\begin{align*}
  W^{*}_{j} = & \argmin_{W_j} -\frac{1}{N}\sum\limits_{i=1}^N  pr_j*q_{i,j}\log(p_{j}(x_i)) + \\ & \hspace{2em}
  (1-q_{i,j})\log(1 - p_{j}(x_i))
    \numberthis \label{eq:training_optimization}
\end{align*}

\noindent where $pr_j$ is the precision constant, a hyperparameter tunes the trade-off between the recall and precision of the classifier. Lower $pr_{j}$ implies higher precision and lower recall. As in Figure \ref{fig:selector_jpeg_hossam_sep}, $p_{i}(x_i)$ is the output of sigmoid function ($\sigma ( . )$) and has two forms:

\begin{align*}
    \text{Form One:} & \hspace{0.5em} p_{j}(x_i) = \sigma(h_{j}(FE(x_i);W_j^{{T}}))
    \numberthis \label{eq:predicted_qf_label_nn_form1} \\
    \text{Form Two:} & \hspace{0.5em} p_{j}(x_i) = \sigma( h_{j}(x_i; W_j^{{T}}))
    \numberthis \label{eq:predicted_qf_label_nn_form2}
\end{align*}

From either (\ref{eq:predicted_qf_label_nn_form1}) or (\ref{eq:predicted_qf_label_nn_form2}), the predicted ON/OFF for a $QF_j$ for each image $x_i$ denoted by $y_j(x_i)$ is:
\begin{equation}
    y_{j}(x_i) =
     \begin{cases}
       \text{1} & \hspace{1em} p_{j}(x_i) \geq DT_j \\
       \text{0} &\quad\text{otherwise} \\
     \end{cases}
     \label{eq:classification_distance_pred}
\end{equation}

\noindent where $DT_j$ is another hyperparameter called the decision threshold for $QF_j$. Higher $DT_{j}$ implies higher precision and lower recall for $y_{j}$. At inference time, Deep Selector-JPEG finally selects the least feasible QF for JPEG compression.

\subsection{Training Description}

We trained the proposed architectures in TensorFlow via stochastic gradient descent. We utilized multi-GPU training via two NVIDIA GeForce RTX 2080 Ti GPUs with a batch size of 100 for two epochs of 20000 steps. The pre-trained DNN classifier for labeling, $S$, is set to be IV3 or MobileNetV2 to represent both heavy-weight and light-weight architectures, respectively. Precision constants and decision thresholds ranged from 0.2 to 0.7 and from 0.5 to 0.9, respectively. 
For the Form One DNN architecture of Deep Selector-JPEG, we produced two selectors, namely \textit{IV3-TwoLayers} and \textit{MobileNet-TwoLayers}, with the underlying DNN $S'$ to be IV3 and MobileNet V2, respectively. In each of these selectors, we froze all layers except for the last two blocks. For the Form two DNN architecture of Deep Selector-JPEG, the entire MobileNetV2 architecture is trained as $S'$ for each binary classification forming \textit{MobileNet-Full}. IV3 selector is trained with labels from IV3, while MobileNetV2 selectors are trained with labels from MobileNetV2.


\begin{figure*}[!ht]
\setlength{\belowcaptionskip}{-3ex}
  \begin{minipage}{\textwidth}
  \centering
  \hspace{0.15\textwidth}
  \hfil \medskip \\[-10ex]
    \setcounter{subfigure}{0}
    \subfloat[GeneralPSNR26Top5Gp9 \label{fig:general_psnr26_top5_gp9}][Top to Bottom: IV3, MobileNetV2]{\includegraphics[width=.4\textwidth, height=0.25\linewidth]{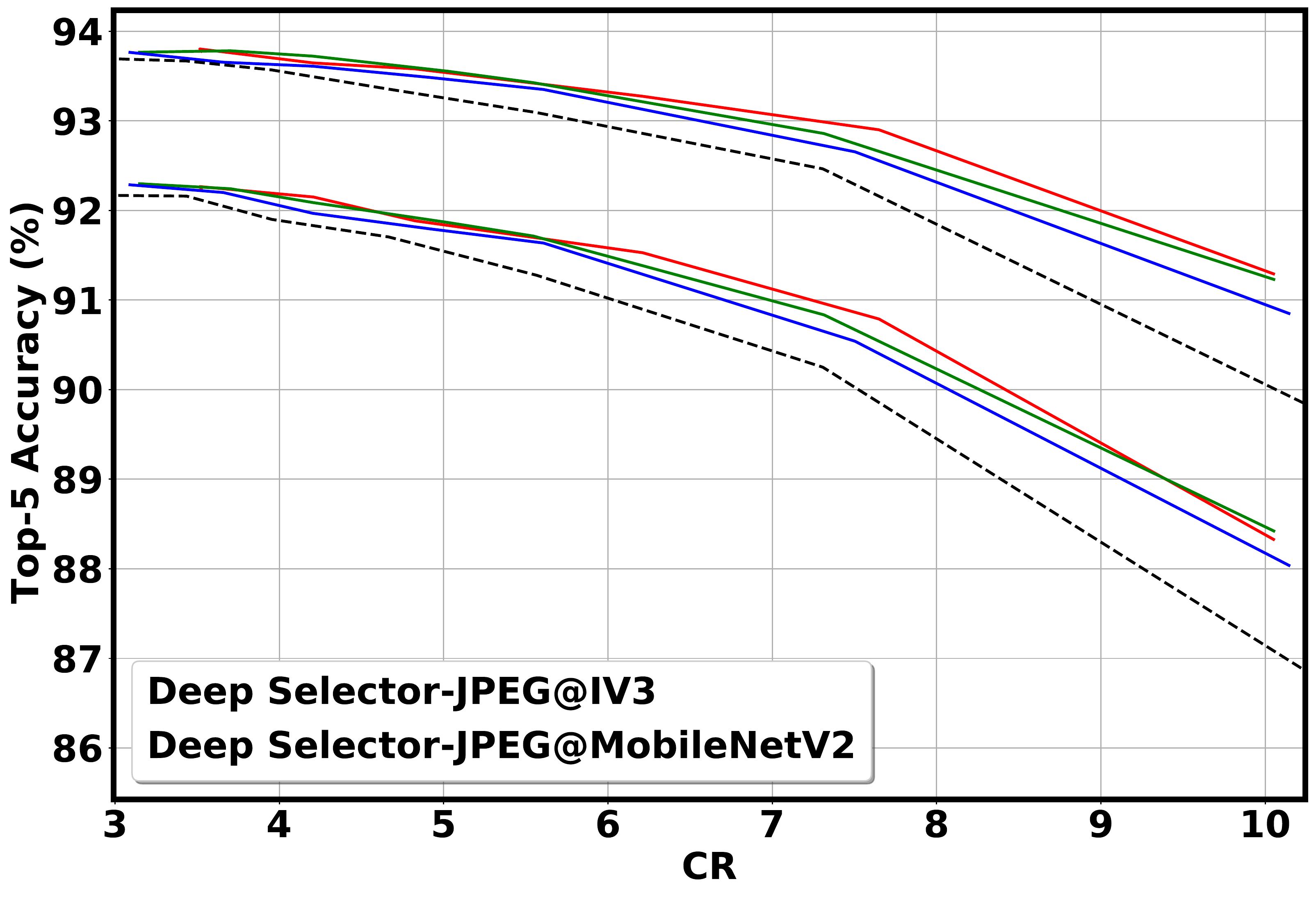}}
    \subfloat[GeneralPSNR26Top5Gp7 \label{fig:general_psnr26_top5_gp7}][Top to Bottom: Pnasnet Large, InceptionResNetV2, Nasnet]{\includegraphics[width=.4\textwidth, height=0.25\linewidth]{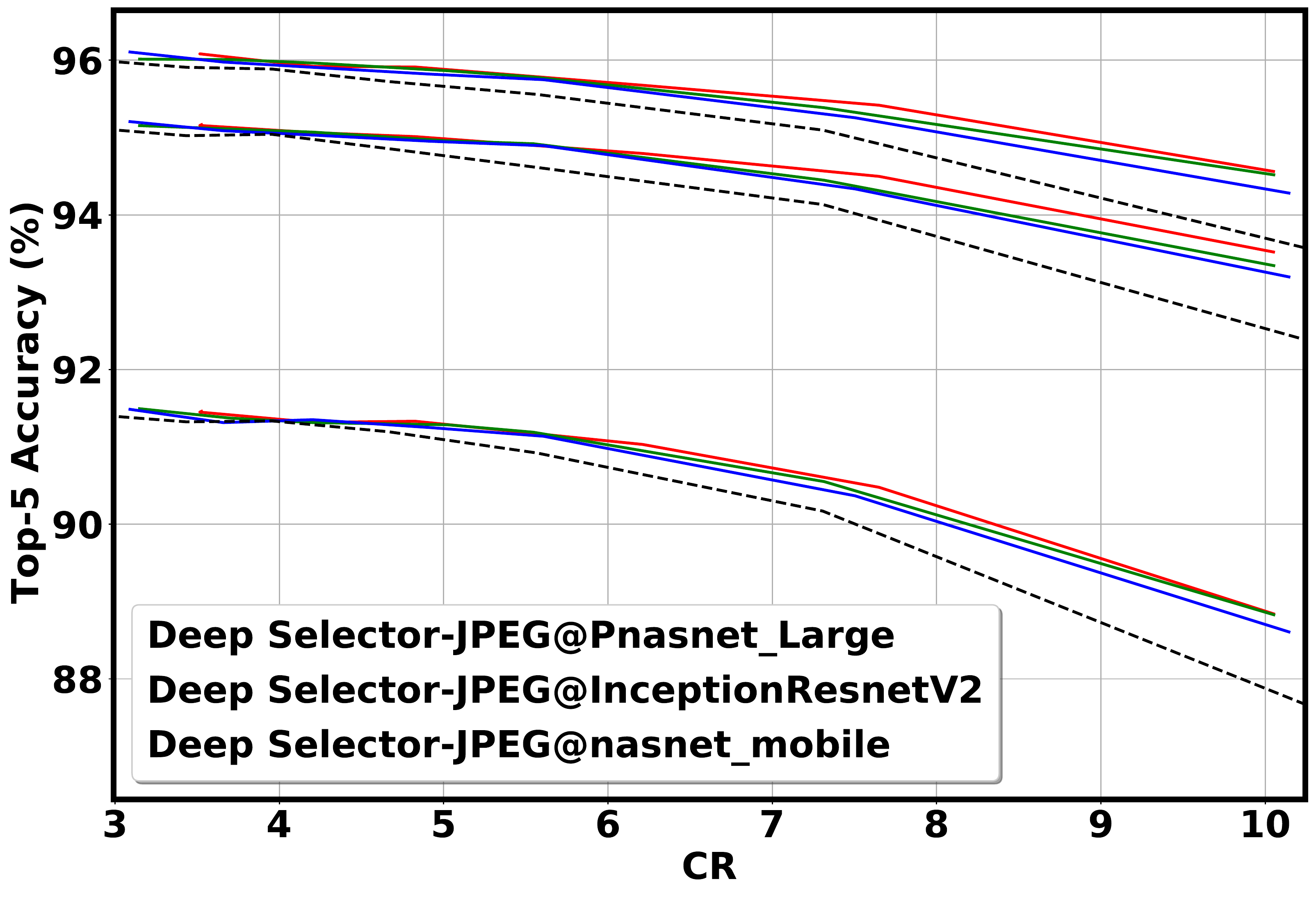}}
  \\
  \subfloat[GeneralPSNR26Top5Gp5 \label{fig:general_psnr26_top5_gp5}][Top to Bottom: IV4, ResNetV2-101, MobileNetV1]{\includegraphics[width=.4\textwidth, height=0.25\linewidth]{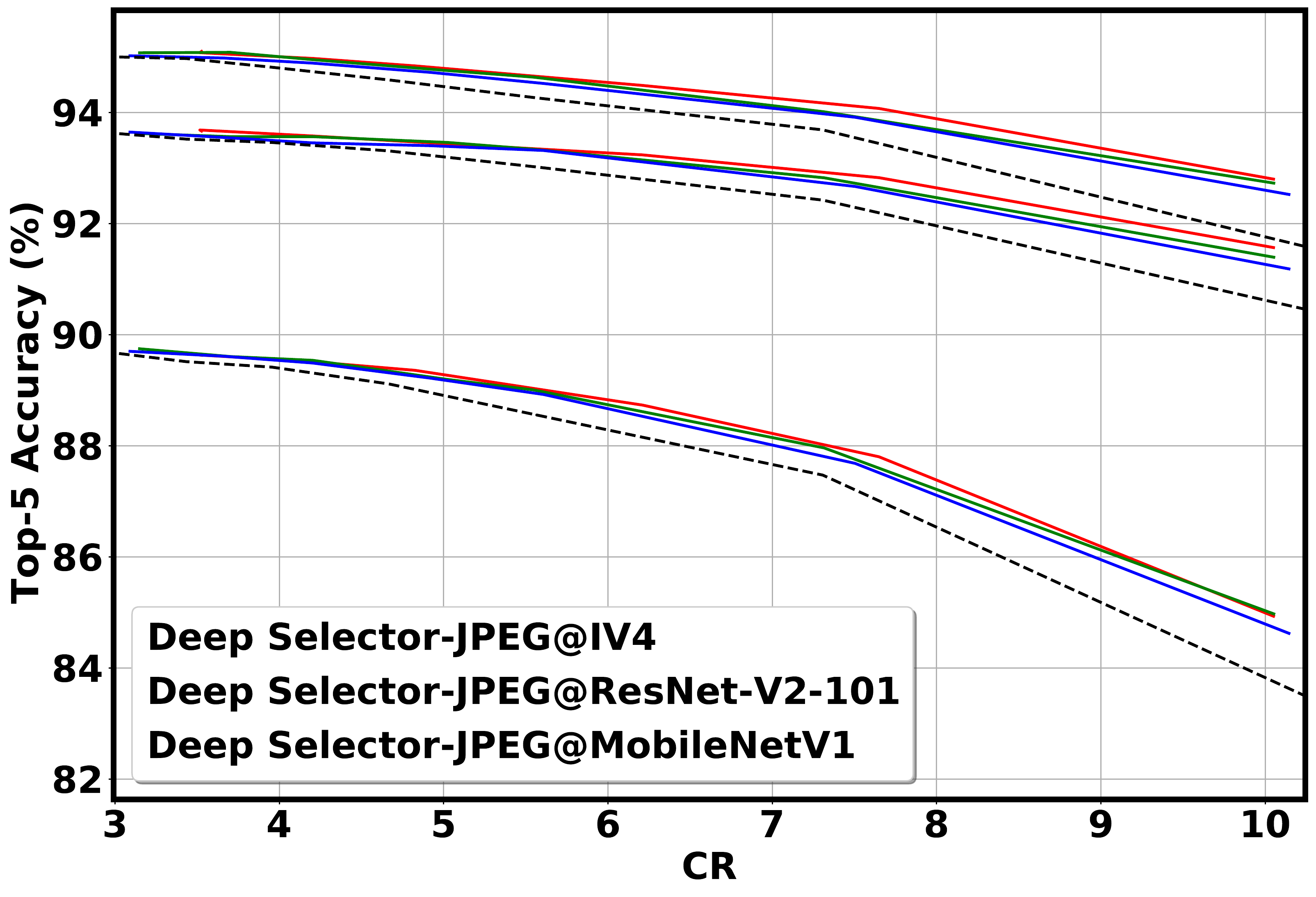}}
  \subfloat[GeneralPSNR26Top5Gp6 \label{fig:general_psnr26_top5_gp6}][Top to Bottom: ResNetV2-50, IV1]{\includegraphics[width=.4\textwidth, height=0.25\linewidth]{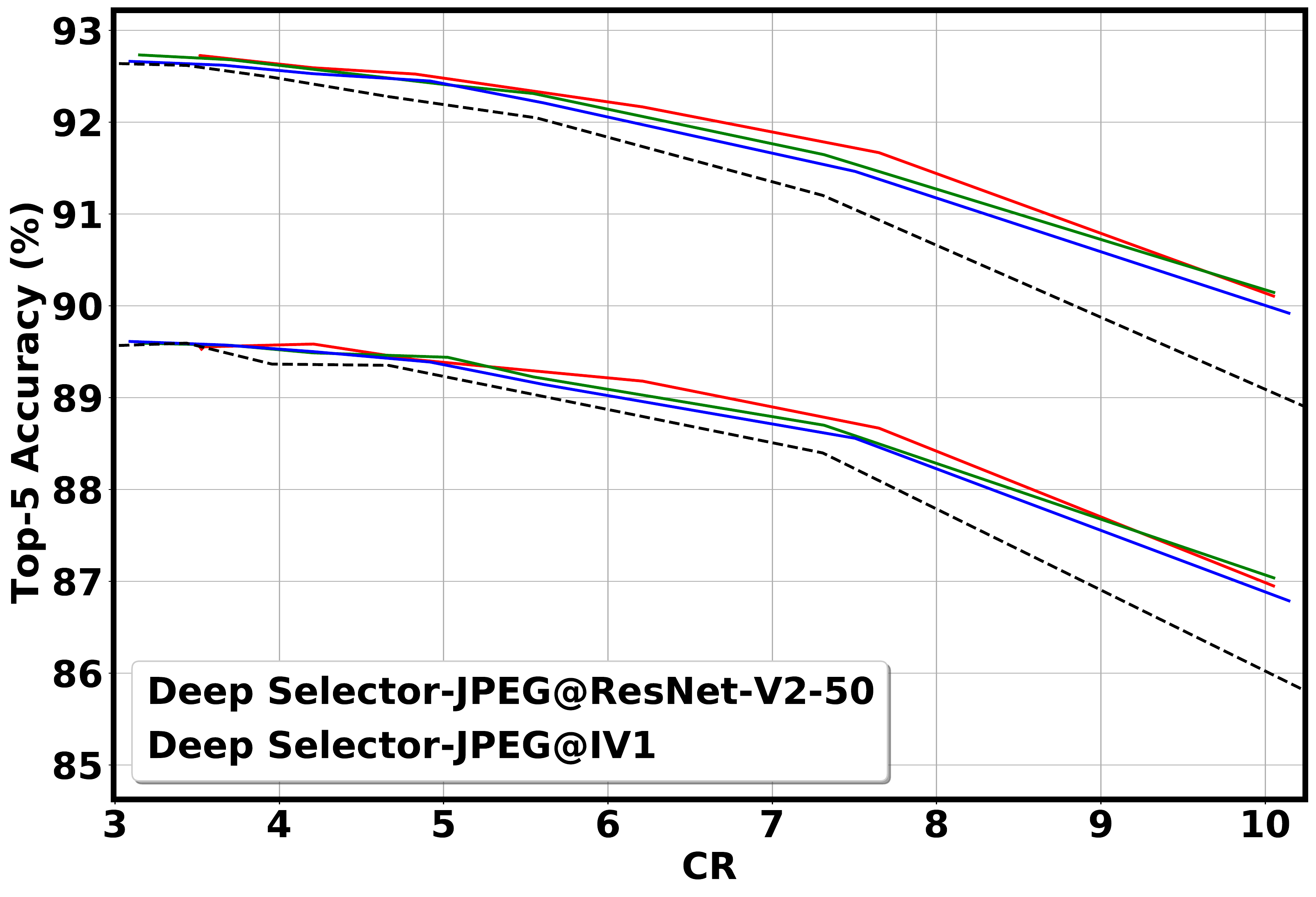}}
  \end{minipage}
  \caption{CR vs Top-5 accuracy of Deep Selector-JPEG when applied to all test DNNs; MS-SSIM $\geq$ 0.8 (PSNR $\geq$ 26dB). The dotted black curve represents JPEG, while the red, green, and blue curves represent IV3-TwoLayers, MobileNet-Full, and MobileNet-TwoLayers, respectively. The legend inside each subfigure outlines the order of test DNNs from top to bottom. For example, the top set of RA curves in Figure \ref{fig:generality_results}(d) is for ResNet-50 V2, and the bottom set of RA curves is for IV1.}
  \label{fig:generality_results}
\end{figure*}

\section{Experimental Results}
\label{sec:results}

This section evaluates the RA performance of Deep Selector-JPEG w.r.t JPEG at the same CR over the entire ImageNet validation set. The RA curves of JPEG are produced by plotting the CR and classification accuracy of test DNNs on 9 compressed clusters of data, each created by compressing all images inside the ImageNet validation set at QFs in \{10,20,..90\}. CR numbers are reported with respect to the high-quality original image in the Imagenet validation set. With taking the entire ImageNet validation set as input stimuli to our experiments, we considered human vision in Section \ref{sec:ra_dnn_known} by carrying out 4 sub-experiments to ensure that the output compressed dataset meets one of four constraints: MS-SSIM $\ge$ 0.8, or 0.85, or 0.9, or 0.95 with 90\% hit rate. The four MS-SSIM constraints lead to ensuring that the PSNR of each output compressed image is greater than 26dB, 28dB, 30dB, and 32dB, respectively. For these MS-SSIM constraints, the starting value of QF in $L$ is set to be 10, 20, 40, and 60 respectively. Section \ref{sec:ra_dnn_known} demonstrates the consistency of Deep Selector-JPEG's RA performance across different DNNs and compares our results with DeepN-JPEG \cite{liu2018deepn}. In Section \ref{sec:selector_jpeg_complexity}, we discuss the complexity of Selector-JPEG.


\subsection{Rate-Accuracy Results With Human Vision Criteria, Consistency, and Comparison with other methods}
\label{sec:ra_dnn_known}

The RA performance of Deep Selector-JPEG using either IV3-TwoLayers, MobileNet-Full, or MobileNet-TwoLayers selectors was evaluated by applying their QF selections at different CRs to 10 popular DNN classifiers with different levels of classification performance on the ImageNet dataset. Figure \ref{fig:generality_results} shows the accuracy gains with respect to JPEG at MS-SSIM $\ge$ 0.8 (PSNR $\geq$ 26dB) with a 90\% hit rate. At CR=3.5x, we observe a slight accuracy gain of approximately 0.2\% relative to JPEG. This gain gradually increases to approximately 1\% when we reach CR=10.05x. It is worth noting that Deep Selector-JPEG achieves 0.9\% top-5 accuracy gain using IV3-TwoLayers at CR=10.05x when applied to a top-performing DNN classifier, Pnasnet\_Large (Baseline's top-5 accuracy=93\% at CR=10.05x). In addition, this figure indicates consistent RA performance gains at MS-SSIM $\geq$ 0.8 across different test DNNs. Due to MobileNet-Full's high capacity, we can observe that MobileNet-Full achieves slightly more accuracy gains than MobileNet-TwoLayers. We can also notice that the performance gain of Deep Selector-JPEG increases slightly when the selector trained via the labeling of the DNN, $S$, is applied to the same DNN classifier compared to other DNNs. In these experiments, we also observed that the original classification accuracy can be maintained at higher CRs.

Deep Selector-JPEG also shows consistent gains at MS-SSIM $\ge$ 0.85 (PSNR $\geq$ 28dB) and CR=6.9x of 0.3\% in terms of top-5 accuracy. In addition, gains of up to 0.1\% at CR=4.4x in terms of accuracy can be achieved when MS-SSIM $\ge$ 0.9 (PSNR $\geq$ 30dB) due to the limited options for our range of QFs in this case, which also exists when MS-SSIM $\geq$ 0.95 (PSNR $\geq$ 32dB). In the future, we hope to adaptively select the QF from a bigger range to give more options to Deep Selector-JPEG, which may further impact the accuracy gains.

Relative to DeepN-JPEG \cite{liu2018deepn}, Deep Selector-JPEG achieves top-5 accuracy near to the original accuracy of the DNN classifiers at CR=3.5x, which is similar to DeepN-JPEG (see the accuracy in Figure \ref{fig:generality_results} vs the original accuracy of the 10 DNNs under test). This similar result is achieved only by adaptively selecting a single parameter QF. We expect that more gains can be achieved by adaptively selecting the whole quantization table for each image as DeepN-JPEG, which is left for future work. It is worth noting that Deep Selector-JPEG improves the overall RA trade-off of JPEG over different DNNs by adaptively selecting a QF for each input image such that the CR and MS-SSIM constraints are satisfied.

\subsection{Time and Complexity Analysis}
\label{sec:selector_jpeg_complexity}

Deep Selector-JPEG's complexity is evaluated in terms of both Multiply-Accumulates (MACs), a popular DNN complexity metric, and total wall clock time. The total wall clock time includes the selector inference as well as JPEG encoding time. With 9 QF selections, the number of MACs of IV3-TwoLayers is 8819.2M, and is 4869.9M for MobileNet-Full. MobileNet-TwoLayers's MACs are 541.5M. From compressing the whole validation set at CR=10.05x and MS-SSIM $\ge$ 0.8 (PSNR $\geq$ 26dB), MobileNet-TwoLayers, MobileNet-Full, and IV3-TwoLayers insignificantly increase the JPEG's time by 2\%, 4.3\%, and 3.74\%, respectively. Other CRs and HV constraints show similar time results. MobileNet-TwoLayers selector has relatively low complexity and its RA performance results are comparable to other proposed selectors, which enables its deployment on edge devices.

\section{Conclusion}

This paper presents Deep Selector-JPEG, an adaptive JPEG compression method that serves image classification while satisfying human vision criteria. Deep Selector-JPEG is designed, based on either a light or heavy-weighted DNN architecture. For each original image,  Deep Selector-JPEG selects adaptively a quality factor to compress the image with its gains in classification accuracy at the same compression ratios between 0.2\% and $\approx$1\% in comparison with JPEG at the same CR. Also, Deep Selector-JPEG can maintain the original classification accuracy at higher CRs. The underlying adaptive selection idea of Deep Selector-JPEG can be transferred to other codecs that can create images at different quality levels such as HEVC \cite{hevcBook}, VVC \cite{wieckowski2019fast}, and deep learning-based codecs. Thus, another extension of this paper is to apply the selection idea to other codecs and computer vision tasks.


\bibliographystyle{IEEEtran}
\bibliography{refs}

\end{document}